\documentclass[aps,prd,twocolumn,groupedaddress,nofootinbib]{revtex4}
\usepackage{bm}
\usepackage{graphicx}
\usepackage{times}
\usepackage{amssymb,amsmath}
\usepackage{rotating}
\usepackage{float}
\usepackage[position=b]{subfig}

\def\araa{Annu. Rev. Astron. Astrophys.}
\def\aap{Astron. Astrophys.}
\def\mnras{Mon. Not. R. Astron. Soc.}
\def\jcap{J. Cosmol. Astropart. Phys.}

\def\apj{Astrophys. J.}
\def\apjl{Astrophys. J. Lett.}
\def\prl{Phys. Rev. Lett.}
\def\prd{Phys. Rev. D}
\def\jhep{J. High Energy Phys.}

\def\beq{\begin{equation}}
\def\eeq{\end{equation}}
\def\bey{\begin{eqnarray}}
\def\eey{\end{eqnarray}}
\def\bfig{\begin{figure}}
\def\efig{\end{figure}}

\def\sun{\odot}
\def\lsim{\mathrel{\raise.3ex\hbox{$<$\kern-.75em\lower1ex\hbox{$\sim$}}}}
\def\gsim{\mathrel{\raise.3ex\hbox{$>$\kern-.75em\lower1ex\hbox{$\sim$}}}}

\begin{document}
\title{Cluster-void degeneracy breaking: Neutrino properties and dark energy}

\author{Martin Sahl\'en$^{1}$}
\email{msahlen@msahlen.net}
\affiliation{Department of Physics and Astronomy, Uppsala University, SE-751 20 Uppsala, Sweden$^{1}$
}

\begin{abstract}
Future large-scale spectroscopic astronomical surveys, e.g.~{\it Euclid}, will enable the compilation of vast new catalogues of clusters and voids in the galaxy distribution. By combining the constraining power of both cluster and void number counts, such surveys could place stringent simultaneous limits on the sum of neutrino masses $M_\nu$ and the dark energy equation of state $w(z) = w_0 + w_a z/(1+z)$. For minimal normal-hierarchy neutrino masses, we forecast that {\it Euclid} clusters + voids ideally could reach uncertainties $\sigma(M_\nu) \lesssim 15$ meV, $\sigma(w_0) \lesssim~0.02$, $\sigma(w_a) \lesssim 0.07$,  independent of other data. Such precision is competitive with expectations for e.g. galaxy clustering and weak lensing in future cosmological surveys, and could reject an inverted neutrino mass hierarchy at $\gtrsim 99\%$ confidence. 
\end{abstract}
\maketitle

\section{Introduction}
Galaxy clusters and voids can be used to place constraints on cosmological models. The abundances of clusters and voids are sensitive to dark energy \cite{2011ARA&A..49..409A, 2015arXiv150307690P,2016ApJ...820L...7S}, modified gravity \cite{2011ARA&A..49..409A,2015MNRAS.450.3319L,2018PhRvD..97j3504S}, neutrino properties \cite{2010JCAP...09..014B,2015arXiv150603088M}, and non-Gaussianity \cite{2010ApJ...724..285C}. 

The ongoing development of void cosmology is a promising new prospect for large-scale astronomical surveys with unprecedented area, depth and resolution. In a series of papers, we are outlining the potential of void surveys to constrain cosmological models, especially in combination with galaxy cluster surveys. In an earlier work \cite{2016ApJ...820L...7S}, we derived the first  cosmological parameter constraints from voids, showing that the joint existence of the largest known cluster and void strongly requires dark energy in the flat $\Lambda$CDM model [with cosmological constant $\Lambda$ and cold dark matter (CDM)]. We also reported a powerful complementarity between clusters and voids in parameter constraints for the $\Lambda$CDM model. In subsequent work \cite{2018PhRvD..97j3504S}, we investigated the complementarity between cluster and void abundances for constraining deviations from general relativity (GR) on cosmological scales. 

Here, we investigate the ability of future {\it Euclid}-like surveys of clusters and voids to constrain neutrino masses and dark energy properties.

\section{Model}
\label{sec:modelmain}

\subsection{Cosmological model}
We assume a flat CDM cosmology, and dark energy with equation of state $w(z)=w_0+w_az/(1+z)=w_0 + w_a(1-a)$, the CPL parameterization \citep{2000ApJ...538..473L,2003PhRvL..90i1301L} (where $a$ is the scale factor). In the following, we will consistently use the term ``dark matter'' to denote all forms of dark matter (including neutrinos), and ``cold dark matter'' for non-neutrino (cold) dark matter only.
The primordial density perturbations are adiabatic and follow a power-law power spectrum. The main fiducial model is specified by the {\it Planck} 2015\footnote{The preprint of the {\it Planck} 2018 results \cite{2018arXiv180706209P} appeared after submission of this work. We focus on the 2015 best-fit cosmology for direct comparability to other forecasts in the literature, but discuss the impact of an alternative fiducial cosmology, with predicted number counts almost identical to those for the {\it Planck} 2018 results (see main text).} \cite{2016A&A...594A..13P} best-fitting flat $\Lambda$CDM parameter values: current Hubble parameter $h=0.673$, current mean matter density $\Omega_{\rm m}=0.314$, dark energy equation of state parameters $w_0=-1, w_a=0$, current mean baryonic matter density $\Omega_{\rm b}=0.0492$, current matter power spectrum normalization $\sigma_8=0.831$, scalar spectral index $n_{\rm s}=0.965$. Neutrinos are modeled with one massive eigenstate and two massless ones, using a sum of neutrino masses $M_{\nu} = 0.06\,{\rm eV}$ (the approximate minimum value allowed by neutrino oscillation data \cite[e.g.][]{2016ChPhC..40j0001P} and the standard value assumed in many cosmological analyses, e.g. for {\it Planck} \cite{2016A&A...594A..13P,2018arXiv180706209P}). Hence, the early-universe effective relativistic degrees of freedom $N_{\rm eff} = 3.046$ \cite{2005NuPhB.729..221M,2016JCAP...07..051D}. We also investigate the sensitivity of our results to the choice of fiducial model, by also considering an alternative fiducial model for which instead $h=0.7, \Omega_{\rm m}=0.3, \sigma_8=0.8$.

\subsection{Surveys}
\label{sec:surv}

We examine the {\it Euclid} Wide Survey \cite{2011arXiv1110.3193L}, covering 15~000 square degrees.   
For clusters, we consider two cases: the full redshift range $z=0.2 - 2.0$ (data set EC), and a low-redshift version covering $z=0.2 - 0.7$ (data set EC-LO). The latter will be used to assess the impact of neglecting cluster-void correlations in the analysis. The lower limit for the cluster mass is $M_{\rm 200,c} = 8 \times 10^{13}\,h^{-1} M_{\sun}$ (where $M_{\rm 200,c}$ is the halo mass within a volume defined by an overdensity threshold of 200 above the critical density). A constant $80\%$ completeness is assumed \cite{2015arXiv150502165S}. We use bins in redshift $\Delta z = 0.1$ and in cluster mass $\Delta \log (M_{\rm 200}) = 0.2$.

For voids, we limit the analysis to spectroscopic data, $z~=~0.7~-~2.0$, to minimize the impact of redshift-space systematics.
Void selection is assumed complete above the limiting radius $R_{\rm lim}(z) = 2 \bar{n}_{\rm gal}^{-1/3}(z)$ \cite{2015arXiv150307690P}, where $\bar{n}_{\rm gal}(z)$ is the mean comoving galaxy number density. See \cite{2018PhRvD..97j3504S} for details. We model the {\it Euclid} galaxy bias as \citep{2016JCAP...05..009R}
\begin{eqnarray}
b_{\rm g}(z) = \sqrt{1+z}\,.
\end{eqnarray}
We note that the galaxy bias in reality depends on the neutrino mass in a scale-dependent manner. The effect on the galaxy bias of an increase in the neutrino mass is a constant enhancement of small-scale bias, and a scale-dependent, increasing bias towards large scales ($k\lesssim 0.1h\,$Mpc$^{-1}$). For the values of neutrino masses we consider, this effect is of the order a few percent \cite[e.g.][]{2014PhRvD..90h3530L, 2016JCAP...11..015B,2018arXiv180807464K}, and we ignore it in our analysis. However, for sufficiently large neutrino masses, and voids defined by highly biased tracers, this effect on the tracer bias can even reverse the pattern of enhancement and suppression of abundances seen in the dark matter field as neutrino mass increases \cite{2018arXiv180807464K}.

For {\it Euclid} voids we consider two binning schemes: 
\begin{enumerate}
    \item[EV-A]{Bins in redshift $\Delta z = 0.1$ and one bin in radius $R>R_{\rm lim}(z)$ and void galaxy field density contrast $\delta^{\rm v}_{\rm g} < -0.8$.}
    \item[EV-B]{Bins in redshift $\Delta z = 0.1$, in void radius $\Delta \log (R)~=~0.1$, and three bins in void galaxy field density contrast in the range $-1 < \delta^{\rm v}_{\rm g}<-0.25$, with $\Delta \delta^{\rm v}_{\rm g} = 0.25$ (corresponding to deep, medium and shallow voids).} 
\end{enumerate}
These binnings should accommodate expected measurement uncertainties. 
The two binning cases can be regarded as worst-case and best-case scenarios with respect to the capability to successfully model and observationally extract void abundances from galaxy surveys.

\subsection{Cluster and void abundance with neutrinos}
\label{sec:model}

We predict cluster and void abundances adopting models and methodology developed in earlier work \cite{2018PhRvD..97j3504S,2016ApJ...820L...7S, Sahlen2009}. As in \cite{2018PhRvD..97j3504S}, we include scatter in cluster mass and void radius determinations, and also vary the characteristic void density contrast (through the parameter $D_{\rm v}$, see below). We neglect cluster-void correlations, but make conservative overestimates of their impact on the results, by considering the low-redshift cluster survey EC-LO which has no overlapping volume with the void surveys. 

In comparison to massless neutrinos, massive neutrinos effectively shift the turnover scale in the matter power spectrum, and suppress power below the neutrino free-streaming scale. This tends to delay and suppress the formation of clusters and voids. In our fiducial dark energy model, the neutrino free-streaming scale is given by \cite[e.g.][]{2013neco.book.....L}
\begin{equation}
k_{\rm FS}^{\nu}(z) \approx 0.8\frac{\sqrt{0.686+0.314(1+z)^3}}{(1+z)^2}\left(\frac{M_{\nu}}{1\,{\rm eV}} \right) h\,{\rm Mpc}^{-1}\,.
\end{equation} 
Additionally, the linear growth rate of over- and underdensities is slightly reduced since free-streaming neutrinos lack gravitational backreaction. 

The local neutrino density will also influence the non-linear evolution of clusters and voids. We describe below the modeling of the effects of neutrinos on non-linear structure formation based on good first-order approximations. 

\subsubsection{1. Cluster abundance}
For galaxy clusters, the effect of neutrinos is modeled following \citet{2010JCAP...09..014B}. On cluster scales, neutrinos free-stream and do not participate in gravitational collapse. Cluster masses are accordingly rescaled: $M~=~4\pi R^3_{\rm L}[(1~-~f_{\nu})\rho_{\rm m}+f_{\nu}\rho_{\rm b}]/3$, where $R_{\rm L}$ is the Lagrangian radius corresponding to the cluster, $\rho_{\rm m}$ is the mean matter density, $f_{\nu}~=~[M_{\nu} / 93\,{\rm eV}] / \Omega_{\rm dm}h^2$ is the fraction of the dark matter density $\Omega_{\rm dm}$ in neutrinos, and $\rho_{\rm b}$ is the mean baryon density. 

\subsubsection{2. Void abundance}
For voids, we model the effect of neutrinos by extending the treatment in \citep{2016ApJ...820L...7S}. When neutrinos have nonzero mass, the neutrino density contributes to the dynamical evolution of a void, but does not have a significant density contrast on its own, except for voids larger than the neutrino free-streaming length \citep[e.g.][]{2014PhRvD..90h3530L, 2015arXiv150603088M, 2016JCAP...11..015B}. If an effective fraction $f_{\rm cl}(R, z; f_\nu)$ of matter participates in clustering below the co-moving scale $R$ at redshift $z$, the total void matter density contrast will be
\begin{eqnarray}
\delta^{\rm v}_{\rm m} = f_{\rm cl}(R, z; f_\nu)\delta^{\rm v}_{\rm cdm}\,, 
\end{eqnarray}
where $\delta^{\rm v}_{\rm cdm}$ is the non-neutrino cold dark matter density contrast of the void. We can write the clustering fraction as 
\begin{equation}
f_{\rm cl}(R, z; f_\nu) = 1-f_{\rm dm}f_\nu f_{\rm FS}(R, z; f_\nu)\,.
\end{equation} 
Here $f_{\rm dm} = 1-\Omega_{\rm b}/\Omega_{\rm m}$ (The earlier work \cite{2016ApJ...820L...7S} contained a typographical sign error in Sec.~3.1.5.) is the fraction of matter in dark matter (including neutrinos), and 
\begin{equation}
f_{\rm FS}(R, z; f_\nu) \approx  1-e^{-R^{\nu}_{\rm FS}(z)/R}
\end{equation}
is the effective fraction of neutrinos that do not participate in clustering due to free-streaming below the neutrino free-streaming length $R^{\nu}_{\rm FS} = 2\pi/k^{\nu}_{\rm FS}$. Since galaxies trace the cold dark matter field, we assume that 
\begin{equation}
\label{eq:galbias}
\delta^{\rm v}_{\rm cdm} = b^{-1}_{\rm g}(z)\delta^{\rm v}_{\rm g} \,,
\end{equation}
where $b_{\rm g}$ is the bias relative to the density contrast $\delta^{\rm v}_{\rm g}$ in the galaxy field of the survey.

We model the evolution of individual voids using the spherical expansion model, whereby the under-density evolves as a separate universe embedded in the background \citep{1992ApJ...388..234B, Jennings2013, 2016JCAP...11..015B}. 
When the fraction of dark matter in neutrinos is small, $0~<~f_\nu~\ll~1$, voids evolve approximately as in a $f_\nu = 0$ cosmology but rescaled by $f_{\rm cl}$: the nonlinear matter density contrast $\delta^{{\rm v}}_{\rm m}$ of a void is well approximated by 
\begin{eqnarray}
\delta^{{\rm v}}_{\rm m} = f_{\rm cl}^{-1}(R, z; f_{\nu})\delta^{{\rm v}, 0}_{\rm m}\,,
\end{eqnarray}
where $\delta^{{\rm v}, 0}_{\rm m}$ is the spherical-expansion solution for $f_\nu = 0$. The relation between non-linear and linear void radii is given by
\begin{equation}
\label{eq:rrl}
\frac{R}{R_{\rm L}} = \left(1+\delta^{{\rm v}}_{\rm m}\right)^{-1/3} = \left(1+ f_{\rm cl}(R, z; f_\nu)\delta^{{\rm v}}_{\rm cdm}\right)^{-1/3}\,.
\end{equation}
The relationship between linear and nonlinear density contrast when $f_\nu = 0$ is well approximated by \cite{1994ApJ...427...51B, Jennings2013} 
\begin{eqnarray}
\delta^{{\rm v}, 0}_{\rm lin, m}(\delta^{{\rm v, 0}}_{\rm m}) = c\left[1-\left(1+\delta^{{\rm v, 0}}_{\rm m}\right)^{-1/c}\right]\,,
\end{eqnarray}
with $c=1.594$ (we take the same approach as e.g.~\cite{2017A&A...607A..24R}). 
Hence, for $0~<~f_\nu~\ll~1$, the relationship between linear and nonlinear density contrast is given by
\begin{eqnarray}
\label{eq:linfirst}
\delta^{{\rm v}}_{\rm lin, m}(R, z) & = & f_{\rm cl}^{-1}(R, z; f_\nu)\delta^{{\rm v}, 0}_{\rm lin, m}\left(\delta^{{\rm v}, 0}_{\rm m}\right) \\
\nonumber 
& = & f_{\rm cl}^{-1}(R, z; f_\nu) \times \\
& & \delta^{{\rm v}, 0}_{\rm lin, m}\left(f_{\rm cl}(R, z; f_\nu) \delta^{{\rm v}}_{\rm m}\right) \\
\nonumber 
& = & f_{\rm cl}^{-1}(R, z; f_\nu) \times \\
\label{eq:linfinal}
& & \delta^{{\rm v}, 0}_{\rm lin, m}\left(f^2_{\rm cl}(R, z; f_\nu) \delta^{{\rm v}}_{\rm cdm}\right)\,.
\end{eqnarray}
In practice, the corrections to the spherical-expansion dynamics compared to $f_\nu = 0$ are $\sim 1-2\%$ for 
values of the neutrino mass $M_\nu < 0.15$ eV.

We employ the volume-conserving ``VdN'' void abundance model \citep{Jennings2013}, with the ``1LDB'' multiplicity function (MF) \citep{2011PhRvL.106x1302C}
\begin{equation}
\label{eq:vmf}
    f(\sigma) = \frac{|\delta^{\rm v}_{\rm lin,m}|}{\sigma \sqrt{1+D_{\rm v}}}\sqrt{\frac{2}{\pi}}\exp\left[-\frac{(|\delta^{\rm v}_{\rm lin,m}|+\beta_{\rm v}\sigma^2)^2}{2\sigma^2 (1+D_{\rm v})} \right] \,,
\end{equation}
where $\delta^{\rm v}_{\rm lin,m}$ is given by Eqs.~(\ref{eq:linfinal}) and (\ref{eq:galbias}). The matter-field dispersion $\sigma(R_{\rm L}, z)$ is calculated on the scale $R_{\rm L}$ given by Eq.~(\ref{eq:rrl}), at redshift $z$. 
Based on the findings of \cite{2017PhRvD..95b4018V}, we set the constant $\beta_{\rm v} = 0$ for simplicity. Note that we vary $D_{\rm v}$ alongside the cosmological parameters in our analysis to account for theoretical uncertainty in the void abundance model. To match the $N$-body results for {\it Euclid}-like surveys in \cite{2015arXiv150307690P}, we first normalize the void number density by the relative VdN volume factor
\begin{equation}
    \left[\frac{V(R_{\rm L})}{V(R)}\right]^{-1}_{\rm Pisani} \,,
\end{equation}
appropriate for those simulations, where $V(R) = 4\pi R^3/3$ is the void volume. We set the fiducial value of $D_{\rm v} = 3.38$. This prescription matches the void abundance results in \cite{2015arXiv150307690P} and \cite{2017PhRvD..95b4018V}, taking the relevant cosmological parameters and survey galaxy bias into account. In \cite{2015arXiv150307690P} it is found that $D_{\rm v}$ can be assumed independent of redshift within uncertainties, and~\cite{2017PhRvD..95b4018V} finds that a single value of $D_{\rm v}$ is valid for different $\delta^{\rm v}_{\rm g}$ as long as the galaxy bias $b_{\rm g}$ is taken into account according to Eq.~(\ref{eq:galbias}) when computing $\delta^{\rm v}_{\rm lin,m}$. This is expected since the matter-field dispersion $\sigma(R_{\rm L}) \sim R_{\rm L}^{-\gamma(R_{\rm L})}$, where $\gamma(R_{\rm L})$ is only weakly dependent on $R_{\rm L}$ \cite{1996MNRAS.281..323V}. A relative difference in radius  between the galaxy and dark matter fields therefore approximately corresponds to rescaling $\sigma$ by a constant, which is what $\sqrt{1+D_{\rm v}}$ effectively does.

\section{Method}
\label{sec:meth}
We compute expected parameter constraints using the Fisher matrix method, based on the Poissonian number counts  \cite{2018PhRvD..97j3504S}. The space of nine free parameters is defined by $\{\Omega_{\rm m}, M_{\nu}, w_0, w_a, \sigma_8,  n_{\rm s}, h, \Omega_{\rm b}, D_{\rm v}\}$. We also consider the eight-parameter constant equation of state case, where $w_a = 0$. Cosmological quantities are computed with a modified version of CAMB \cite{2000ApJ...538..473L}. 

We compute forecast Bayes factors for a normal neutrino hierarchy vs. an inverted hierarchy, assuming the normal-hierarchy fiducial model. From this we can estimate the significance with which a minimal normal hierarchy can be distinguished from an inverted hierarchy in the model inference sense, with the different surveys considered. 

The posterior odds for normal hierarchy vs. inverted hierarchy is given by \cite{2012PhRvD..86k3011Q, 2014JHEP...01..139B, 2017PhRvD..96l3503V,deSalas:2017kay}
\begin{eqnarray}
\frac{p({\rm NO} | d)}{p({\rm IO} | d)} = B_{\rm NO, IO}  \frac{\pi({\rm NO})}{\pi({\rm IO})}\,,
\end{eqnarray}
where ``NO'' denotes normal ordering, ``IO'' denotes inverted ordering, $d$ is the data under consideration, $B_{\rm NO, IO}$ is the Bayes factor (see below), and $\pi$ denotes the prior model probabilities. We assume here that $\pi({\rm NO}) = \pi({\rm IO})$, and therefore the posterior odds are given by $B_{\rm NO, IO}$. 
The Bayes factor $B_{\rm NO, IO}$ is the Bayesian evidence ratio:
\begin{eqnarray}
\nonumber 
B_{\rm NO, IO}  = && \\ & \frac{\int_{M_{{\rm NO}}^{\rm min}}^{\infty} \int_\theta \mathcal{L}(d | \theta, M_\nu; {\rm NO}) \pi(\theta)\pi (M_\nu) {\rm d} \theta {\rm d}M_\nu}{\int_{M_{\rm IO}^{\rm min}}^{\infty} \int_\theta \mathcal{L}(d | \theta, M_\nu; {\rm IO}) \pi(\theta)\pi(M_\nu) {\rm d} \theta {\rm d}M_\nu}\,, &
\end{eqnarray}
where $\mathcal{L}(d | \theta, M_\nu; H)$ is the likelihood function for the data $d$ given model parameters $\theta$ and $M_\nu$ under the neutrino hierarchy hypothesis $H$. The priors $\pi(\theta)$ and $\pi(M_\nu)$ are chosen as flat (uniform) in all parameters and nonrestrictive with respect to the likelihood function. 

We set $M_{\rm NO}^{\rm min} = 0.06$ eV, $M_{\rm IO}^{\rm min} = 0.10$ eV \cite{1674-1137-38-9-090001}.
The confidence level of the rejection of the disfavoured model (in this case the inverted neutrino hierarchy, since we are assuming a minimal normal neutrino hierarchy) is
\begin{equation}
1-\alpha = 1 - |B_{\rm NO,IO}|^{-1}
\end{equation}
which can also be translated to an effective ``number of $\sigma$'' confidence level $n_{\sigma}^{\rm eff}$ defined by the equation
\begin{equation}
    {\mathrm{erf}}\left(\frac{n_{\sigma}^{\rm eff}}{\sqrt{2}}\right) = 1-\alpha \,.
\end{equation}
Dark energy Figures of Merit are computed as
\begin{equation}
    {\rm FoM}(w_0, w_a) = \frac{1}{\sqrt{\det {\rm cov}(w_0, w_a)}} \,.
\end{equation}

\begin{figure}[t]
\includegraphics[trim={2.8cm 0.1cm 1.2cm 1.4cm},clip,width=0.5\textwidth]{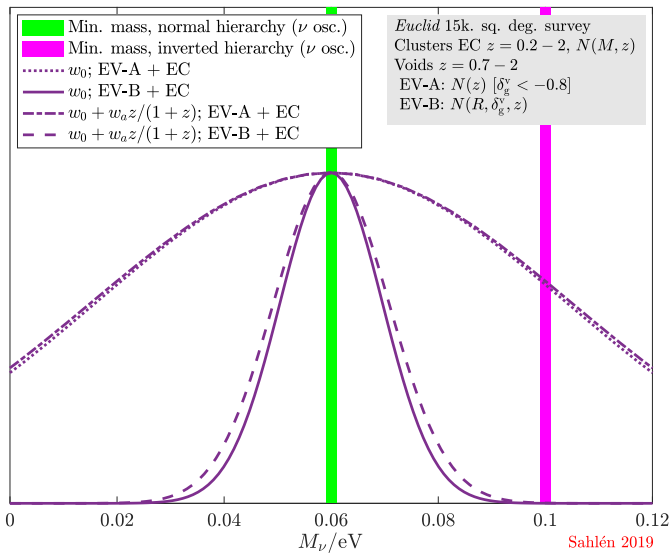}
\caption{ \label{fig:marg} Forecast marginalized probability density functions (pdfs) for the summed neutrino mass $M_{\nu}$ from cluster and void abundances in the {\it Euclid} survey. A constant ($w_0$) or time-varying ($w_0, w_a$) dark energy equation of state is assumed. See 
Secs.~\ref{sec:modelmain} and \ref{sec:meth} for definitions and details.}
%\vspace{-10pt}
\end{figure}

\section{Results}
\subsection{Forecast parameter constraints}

\begin{table*}[htp]
\vspace{-5pt}
\caption{Forecast $68\%$ parameter uncertainties (unless otherwise specified), significance levels of neutrino-hierarchy model inference, and dark energy Figures of Merit from cluster and void abundances in future {\it Euclid}-like surveys. See 
Secs.~\ref{sec:modelmain} and \ref{sec:meth} for definitions and details.}
\label{tab:constr}
\begin{center}
\begin{tabular}[b]{|c|c|c|c|c|c|c|c|c|c|c|c|c|c|c|c|}
\hline 
 & \multicolumn{8}{|c|}{Parameter Inference} & \multicolumn{3}{|c|}{Neutrino Mass Ordering} & \multicolumn{1}{|c|}{DE}
\\ \cline{2-12}
Data set & $\sigma(\Omega_{\rm m})$ & $\sigma(M_{\nu})$ / CL & $\sigma(w_0)$ & $\sigma(w_a)$ & $\sigma(\sigma_8)$ & $\sigma(n_{\rm s})$ & $\sigma(h)$ & $\sigma(\Omega_{\rm b})$ / CL & $\ln \left(B_{\rm NO,IO}\right)$ & Odds $B_{\rm NO,IO}$ & $n_\sigma^{\rm eff}$ & FoM \\
\cline{2-13}
& \multicolumn{12}{|c|}{$w(z) = w_0$} \\
\hline
EV-A & $0.04$ & $<0.4$ eV (95\%)% $0.148$ eV 
& $0.26$ & -- &  $1.7$ & $0.82$ & $0.05$ & $\leq \Omega_{\rm m}$ & $0.3$ & $1.3 : 1$ & $0.27$ & -- \\
EV-B & $0.004$ & $15$ meV  & $0.009$ & -- & $0.11$ & $0.02$ & $0.008$ & $0.005$ & $4.9$ & $130 : 1$ & $2.7$ & --\\ 
EC & $0.002$ & $<1.6$ eV (95\%) %$0.786$ eV 
& $0.007$ & -- & $0.01$ & $0.08$ & $0.03$ & $<0.13$ (95\%) & $0.0$ & $1.0 : 1$ & $0.05$ & --\\
EV-A+EC & $0.0006$ & $<0.18$ eV (95\%) %$44$ meV 
& $0.003$ & -- & $0.001$ & $0.04$ & $0.02$ & $0.02$ & $1.0$ & $2.8 : 1$ & $0.91$ & --\\ 
EV-A+EC-LO & $0.0008$ & $<0.18$ eV (95\%) %$47$ meV 
& $0.01$ & -- & $0.005$ & $0.09$ & $0.02$ & $0.03$ & $0.9$ & $2.5 : 1$ & $0.85$ & -- \\ 
EV-B+EC & $0.0005$ & $10$ meV & $0.003$ & -- & $0.0006$ & $0.01$ & $0.007$ & $0.003$ & $9.7$ & $1.6 \times 10^{4} : 1$ & $3.5$ & -- \\ 
EV-B+EC-LO & $0.0007$ & $11$ meV & $0.003$ & -- & $0.002$ & $0.01$ & $0.007$ & $0.003$ & $8.2$ & $3.6 \times 10^{3} : 1$ & $3.4$ & -- \\ 
\hline
& \multicolumn{12}{|c|}{$w(z) = w_0 + w_a z/(1+z)$} \\
\hline
EV-A & $0.04$ & $<0.8$ eV (95\%)%$0.352$ eV 
& $1.1$ & $4.0$ & $1.8$ & $1.1$ & $0.08$ & $\leq \Omega_{\rm m}$ & $0.1$ & $1.1 : 1$ & $0.11$ & $1$ \\
EV-B & $0.004$ & $17$ meV  & $0.03$ & $0.15$ & $0.11$ & $0.02$ & $0.009$ & $0.005$ & $4.0$ & $54 : 1$ & $2.4$ & $750$ \\ 
EC & $0.003$ & $<1.7$ eV (95\%)%$0.805$ eV 
& $0.01$ & $0.04$ & $0.01$ & $0.08$ & $0.03$ & $<0.14$ (95\%) & $0.0$ & $1.0 : 1$ & $0.05$ & 3500 \\
EV-A+EC & $0.001$ & $<0.18$ eV (95\%) %$45$ meV 
& $0.01$ & $0.04$ & $0.002$ & $0.04$ & $0.02$ & $0.02$ & $1.0$ & $2.7 : 1$ & $0.89$ & 7500 \\ 
EV-A+EC-LO & $0.003$ & $<0.19$ eV (95\%) %$52$ meV 
& $0.03$ & $0.12$ & $0.006$ & $0.10$ & $0.03$ & $0.03$ & $0.8$ & $2.3 : 1$ & $0.77$ & 640 \\ 
EV-B+EC & $0.001$ & $11$ meV & $0.009$ & $0.03$ & $0.001$ & $0.01$ & $0.007$ & $0.004$ & $8.2$ & $3.6 \times 10^{3} : 1$ & $3.4$ & 12000 \\ 
EV-B+EC-LO & $0.002$ & $15$ meV & $0.02$ & $0.07$ & $0.003$ & $0.02$ & $0.008$ & $0.004$ & $4.9$ & $130 : 1$ &  $2.7$ & 4600 \\ 
\hline
\end{tabular}
\end{center}
\vspace{-12pt}
\end{table*}%

\begin{figure*}[t]
\includegraphics[trim={0.6cm 0.4cm 0.1cm 0.1cm},clip,width=\textwidth]{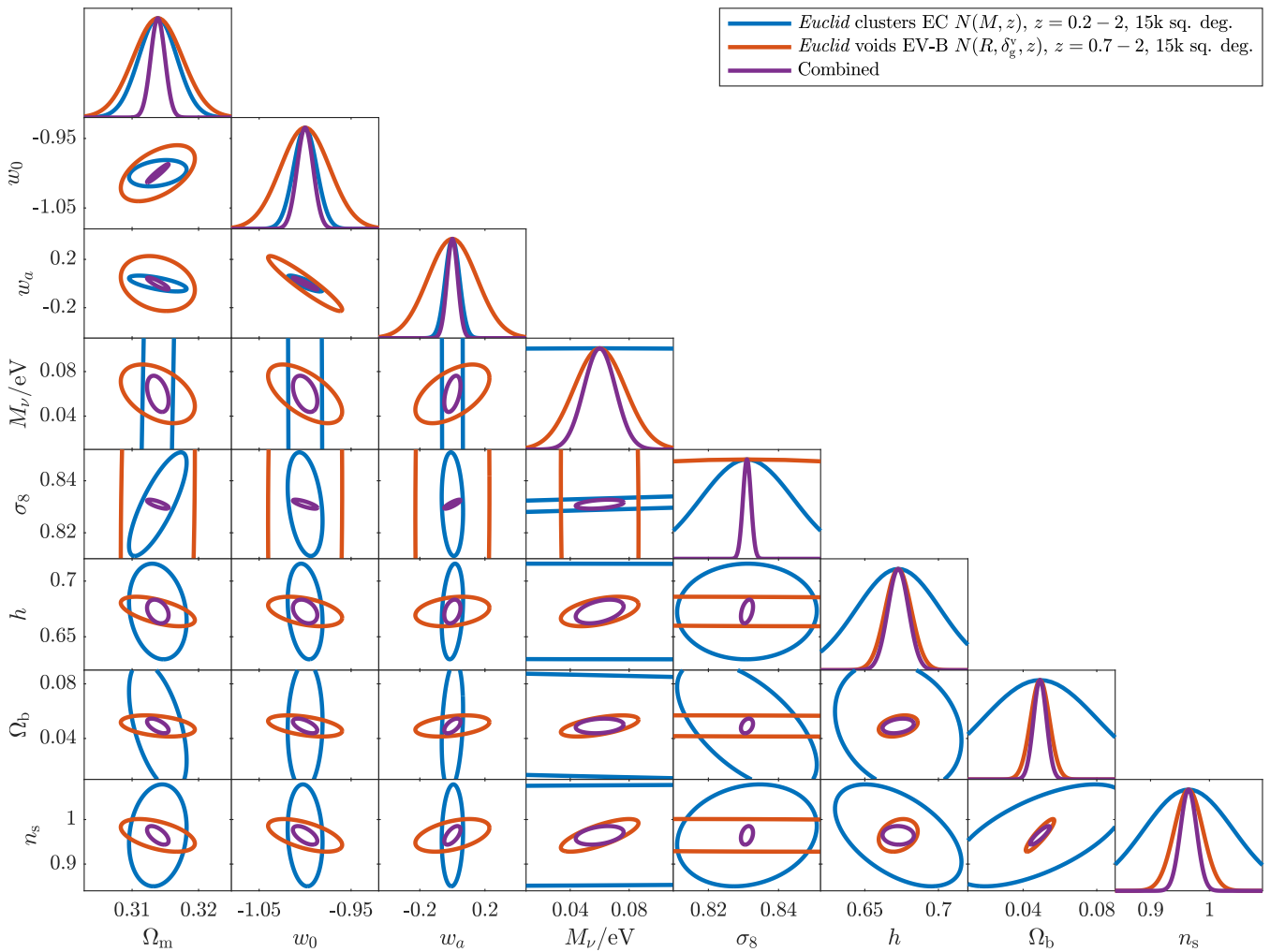}
\caption{ \label{omms8} Forecast $68\%$ parameter contours, and marginal probability density functions, from cluster and void abundances in future {\it Euclid} surveys. A dark energy equation of state $w(z)=w_0+w_az/(1+z)$ is assumed. See Secs.~\ref{sec:modelmain} and \ref{sec:meth} for definitions and details.}
%\vspace{-10pt}
\end{figure*}

The marginalized constraints from {\it Euclid} on the summed neutrino mass $M_{\nu}$ are shown in Fig.~\ref{fig:marg}, while the results for all the cosmological parameters are reported in Table~\ref{tab:constr}. Figure~\ref{omms8} shows the full set of one-dimensional and two-dimensional marginalized parameter constraints for cluster and void counts in the $(w_0, w_a)$ dark energy model, for EV-B and EC. We omit the results on the nuisance parameter $D_{\rm v}$ for brevity. Notably, we see in Fig.~\ref{omms8} that cluster and void constraints are orthogonal among many of the model parameters. Clusters are most sensitive to dark energy parameters, and voids most sensitive to the sum of neutrino masses. The combination of deep and shallow void counts (in EV-B) can powerfully break degeneracies between background expansion, shape of the power spectrum, and growth history \cite{2018PhRvD..97j3504S}, akin to a multi-tracer approach. Clusters of galaxies have complementary sensitivity to expansion history, growth history and power spectrum scales, so that combining cluster and void counts further breaks degeneracies \cite{2016ApJ...820L...7S,2018PhRvD..97j3504S}. In terms of structure growth, a void survey at redshift $z$ can be regarded as roughly equivalent to a cluster survey at redshift $z + 0.5$ (since the nonlinear collapse / shell-crossing thresholds are $\delta_{\rm nl}^{\rm cluster} \sim 1.7$, $\delta_{\rm nl}^{\rm void} \sim -2.7$ in terms of extrapolated linear density), but with orthogonal degeneracy between $\Omega_{\rm m}$ and $\sigma_8$. These features explain the ability of void counts to constrain all parameters in our model, the strengthening of constraints when cluster counts are added, and the fact that many parameter constraints remain nearly unchanged when an additional parameter ($w_a$) is considered.

We see from Table~\ref{tab:constr} that if voids binned in void radius and density contrast (EV-B) are combined with clusters (EC), we expect to measure the summed neutrino mass with an uncertainty $\sigma(M_\nu) \lesssim 15$ meV.
When the total abundances of deep voids above the limiting radius in redshift bins (EV-A) are combined with clusters (EC), we expect only marginally competitive constraints of $M_\nu \lesssim 0.19$~eV ($95\%$~CL). Still, the combined constraints are significantly tighter than for the individual cluster and void surveys, and provide an independent test based on large-scale structure only. These results do not change when a constant (one-parameter) or time-dependent (two-parameter) dark energy equation of state is assumed (see Fig.~\ref{fig:marg}). 

For the dark energy equation of state with EV-B + EC, we forecast combined cluster-void constraints $\sigma(w_0)~\lesssim~0.02, \sigma(w_a) \lesssim 0.07, {\rm FoM}(w_0, w_a) \gtrsim 4600$. In the worst-case scenario EV-A + EC, the combined cluster-void constraints do not improve on the cluster-only (EC) constraints $\sigma(w_0) \lesssim 0.01, \sigma(w_a) \lesssim 0.04, {\rm FoM}(w_0, w_a) \gtrsim 3500$.

For simplicity, we have neglected spatial cluster-void correlations. We may therefore be overestimating the statistical power of our joint cluster-void analysis. To investigate the possible degradation of parameter constraints due to cluster-void correlations, we consider the alternative cluster survey EC-LO truncated at $z=0.7$ (the lower redshift limit of the void survey). Thus, we throw away the clusters in the EV-A/B + EC overlapping volume across $z = 0.7-2.0$. The uncertainties on $M_\nu$ are robust to within $10\%$ in this analysis. The uncertainties on $w_0$ and $w_a$ increase by a factor of a few, but we note that the full cluster-only parameter uncertainties (EC) are significantly smaller in comparison.

To demonstrate that the results are stable against changes in fiducial cosmology, we repeated the analysis using an alternative fiducial model, with differing values $h~=~0.7, \Omega_{\rm m}~=~0.3, \sigma_8=0.8$. We have confirmed that the predicted number counts for this model are identical with those for the {\it Planck} 2018 best-fit model \cite{2018arXiv180706209P}, to within a few per cent. For these two models, the parameter uncertainties are marginally larger for the different individual and joint cases, but within rounding error. The only exception is the case where we consider EV-A data only, for which parameter uncertainties are a factor $2$--$4$ larger (except for $\sigma(\sigma_8)$, which is unchanged).

\begin{figure*}[t]
	\subfloat[Deep voids ($\delta_{\rm g}^{\rm v}= -0.85$).]{\includegraphics[trim={0.9cm 0.6cm 0.4cm 0.1cm},clip,width=0.5\textwidth]{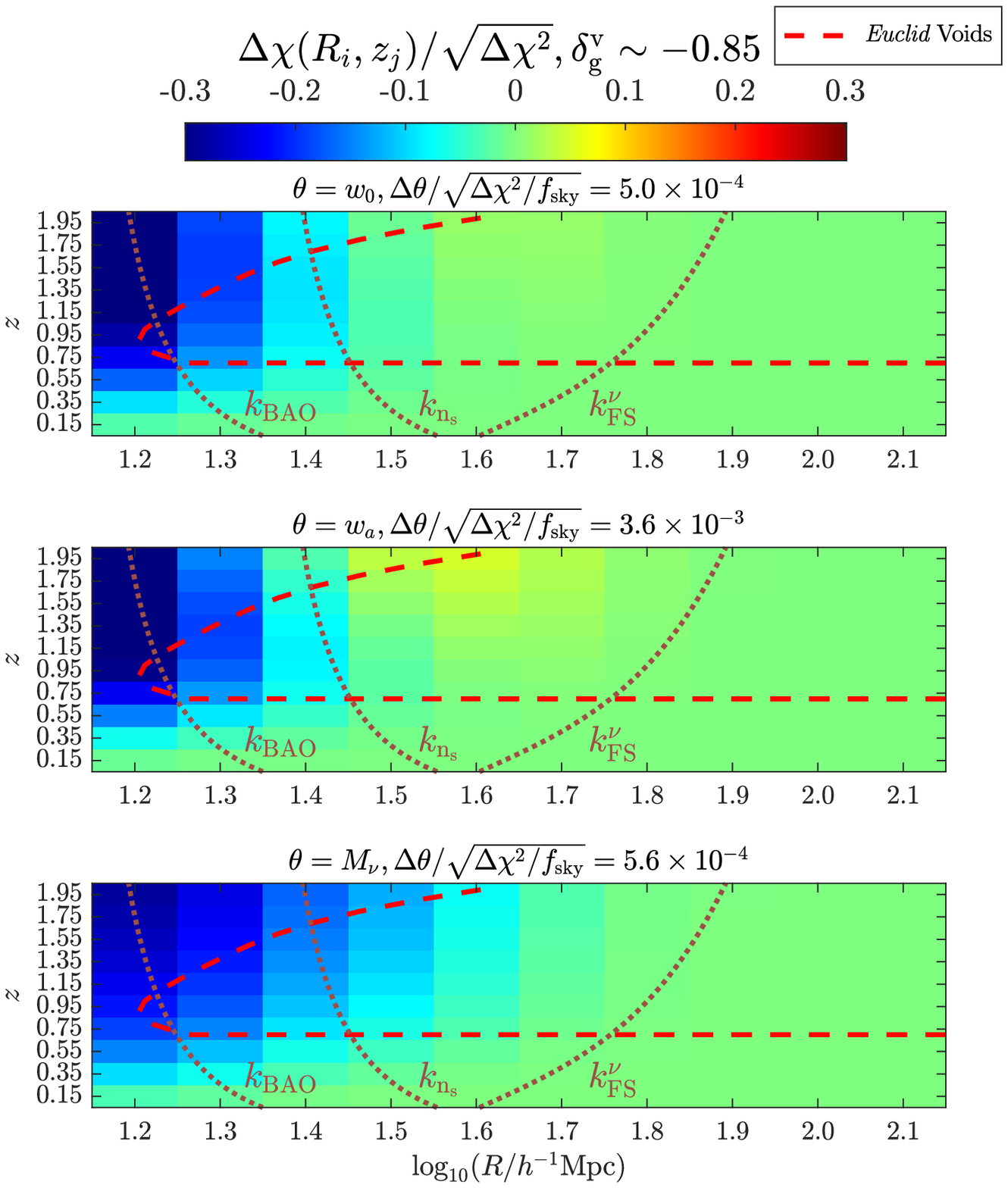}}
	\subfloat[Shallow voids ($\delta_{\rm g}^{\rm v}= -0.25$).]{\includegraphics[trim={0.9cm 0.6cm 0.4cm 0.1cm},clip,width=0.5\textwidth]{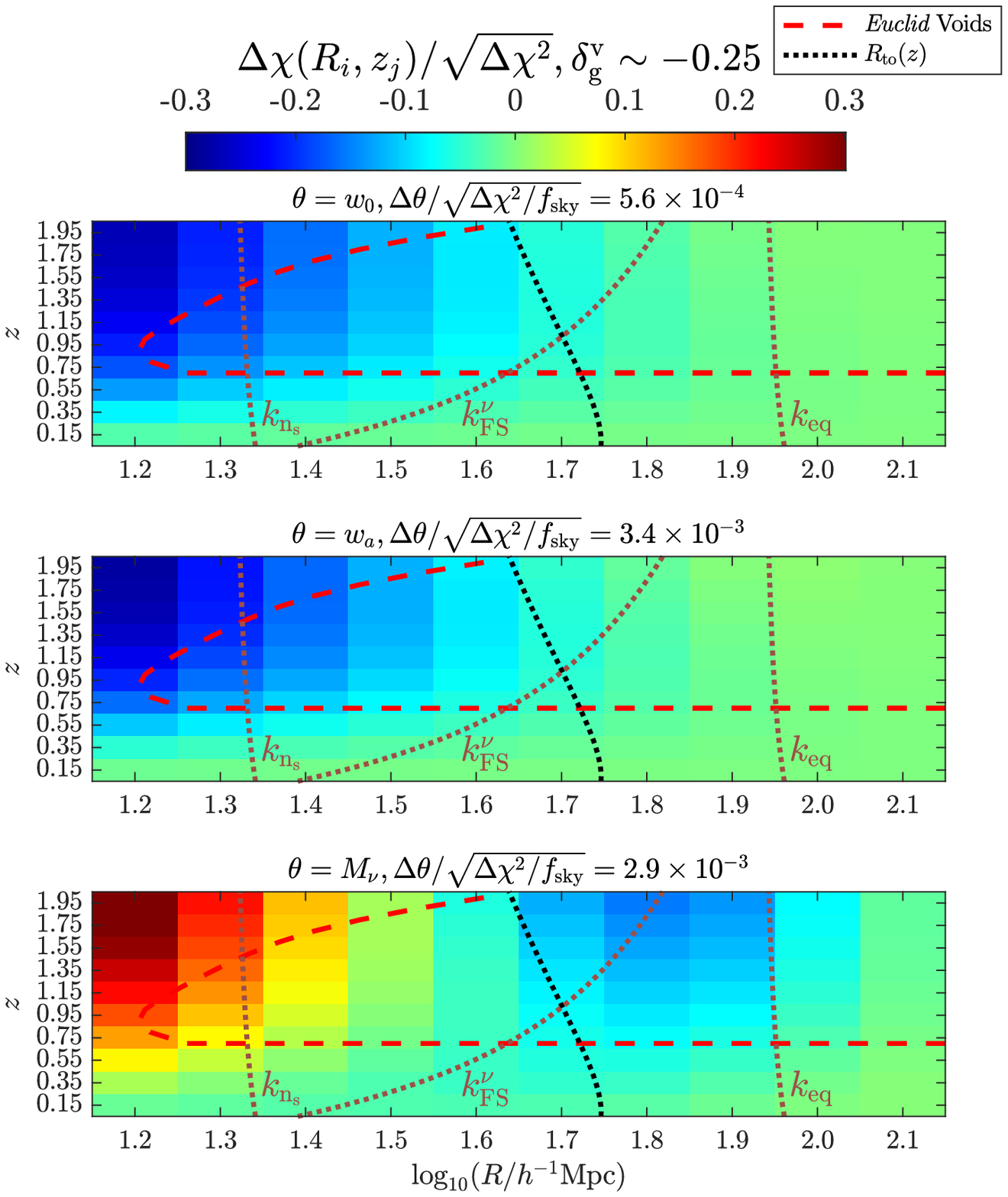}}
%\vspace{-10pt}
\caption{ \label{fig:sens} Void parameter sensitivity. We here use a generic void survey with $R_{\rm lim} = 14\,h^{-1}{\rm Mpc}$ and $z=0.05 - 2.05, \Delta \log_{10} (R/h^{-1}Mpc) = 0.1, \Delta z = 0.2$. For each parameter, the figure shows $\Delta \chi^{\rm rel}_{i,j}$ when that parameter {\it only} is varied. Hence, $\sigma_8$ is kept normalized to the fiducial value when other parameters are varied. See \cite{2018PhRvD..97j3504S} for additional parameters. The turnover radius $R_{\rm to}(z)$ is shown in black, dotted lines. Scales related to the cosmological parameters are shown in brown, dotted lines ($k_{\rm BAO} = 0.06h$ Mpc$^{-1}$, $k_{\rm n_s} = 0.05h$ Mpc$^{-1}$, $k_{\rm FS}^{\nu} \sim 0.003 - 0.05 h$ Mpc$^{-1}$, $k_{\rm eq} = 0.012h$~Mpc$^{-1}$). The coverage of the {\it Euclid} void surveys in terms of limiting radii and redshift is shown in red, dashed lines. See Sec.~\ref{sec:parconst} for definitions and details.}
\end{figure*}

\subsection{Void parameter sensitivity}
\label{sec:parconst}
The sensitivity of void counts to changes in the parameters $w_0$, $w_a$ and $M_\nu$ is shown in Fig.~\ref{fig:sens} (see also Fig.~3 of Ref.~\cite[]{2018PhRvD..97j3504S}). We illustrate sensitivity using the quantity
\begin{eqnarray}
\Delta \chi^{{\rm rel}}_{i,j}(\Delta \theta_k) \equiv \frac{\Delta \chi(R_i, z_j; \Delta \theta_k, f_{\rm sky})}{\sqrt{\Delta \chi^2(\Delta \theta_k, f_{\rm sky})}}\,,
\end{eqnarray}
for a small positive one-parameter shift $\Delta \theta_k$ away from the fiducial cosmological model. Here, the quantity $\Delta \chi$ represents the number count change in bin $i, j$, in units of Poisson uncertainty, under the shift $\Delta \theta_k$:
\begin{equation}
\Delta \chi(R_i, z_j; \Delta \theta_k, f_{\rm sky}) =
\sqrt{\frac{2f_{\rm sky}}{\bar{N}_{i, j}}}\frac{\partial \bar{N}_{i,j}}{\partial \theta_k} \Delta \theta_k \,. \\
\end{equation}
The quantity $\Delta \chi^2$ is the total change in  $\chi^2$ across all bins under the shift $\Delta \theta_k$: 
\begin{equation}
\Delta \chi^2(\Delta \theta_k, f_{\rm sky}) = \Sigma_{i,j} \Delta \chi^2(R_i, z_j; \Delta \theta_k, f_{\rm sky}) \,.
\end{equation}
Above, bins in radius and redshift are indexed by $i$ and $j$, $\bar{N}_{i,j}$ is the fiducial expected number of voids in bin $(i,j)$, and $f_{\rm sky}$ is the survey fractional sky coverage. Note that $\Delta \chi^{{\rm rel}}_{i,j}$ is independent of the survey sky fraction $f_{\rm sky}$, and represents the statistical significance of the number count change in bin $i, j$ relative to average statistical significance of number count changes in all bins, under the shift $\Delta \theta_k$ \cite{2018PhRvD..97j3504S}.

Figure~\ref{fig:sens} also shows the void turnover scale $R_{\rm to}(z)$ (right panel only, as it is below the limiting radius in the left panel), here defined by
\begin{equation}
\nu(R_{\rm to}, z) = 1 \,,
\end{equation}
where 
\begin{equation}
    \nu(R, z) = 
\frac{|\delta^{\rm v}_{\rm lin,m}(R,z)|^2}{\sigma^2(R,z) (1+D_{\rm v})} \,,
\end{equation}
assuming $\beta_{\rm v}=0$ in Eq.~(\ref{eq:vmf}). Above the turnover radius, the sensitivity to the matter power spectrum and growth history gradually dominates over the sensitivity to the expansion history \cite{2018PhRvD..97j3504S}. Note that the linear-density threshold $\delta^{\rm v}_{\rm lin, m}$ is redshift-dependent, as the survey galaxy bias.

The top two panels in Fig.~\ref{fig:sens} show that when the dark energy equation of state ($w_0$ or $w_a$) is increased, the effect on void abundances is simple for both deep ($\delta_{\rm g}^{\rm v} \sim -0.85$) and shallow ($\delta_{\rm g}^{\rm v} \sim -0.25$) voids: the cosmic volume is reduced, and hence the void abundance suppressed. For sufficiently rare voids, however, the relatively enhanced growth overtakes the volume suppression to enhance the void abundance. The impact of variations in the dark energy equation of state and the linear growth rate is discussed in more detail in \cite{2018PhRvD..97j3504S}.

When the neutrino mass $M_\nu$ is varied, the effect is significantly different for deep and shallow voids, as seen in the bottom panels of Fig.~\ref{fig:sens}. Thus, the balance of the abundances of small and large voids of different depth should be a good probe of neutrino mass. A key to this effect is the location of the turnover radius, which implies that  the effect is the result of an interplay between $\sigma(R,z)$, $b_{\rm g}(z)$, $f_{\rm cl}(R,z)$ and $D_{\rm v}$ (in addition to the dependence on $\delta_{\rm g}^{\rm v}$). Hence, the two distinct effects can also be replicated by considering tracers with low and high bias, respectively. 

In the following, we describe the impact on void abundances from changing the total neutrino mass, as demonstrated in Fig.~\ref{fig:sens}. If $M_\nu$ is increased, the matter power spectrum is significantly suppressed below the free-streaming scale $k_{\rm FS}^{\nu}$. The suppression tapers out towards $k_{\rm eq}~\sim~0.01h\,$Mpc$^{-1}$. In Fig.~\ref{fig:sens} we keep the matter power spectrum normalized to the fiducial value of $\sigma_8$ (at $z=0$) when increasing $M_\nu$. Hence, the overall effect is to suppress the matter power spectrum for $k\gtrsim k_{\rm eq}$ and enhance it for $k\lesssim k_{\rm eq}$; the relative difference increasing with redshift.

The value of $\sigma(R,z)$, the matter power spectrum at redshift~$z$ averaged on the scale $R$, is suppressed or enhanced with the power spectrum itself. Since 
\begin{equation}
\label{eq:dlnfv}
\frac{d\ln f}{d \ln \sigma} = \nu
- 1\,,
\end{equation}
for $\beta_{\rm v}=0$ in Eq.~(\ref{eq:vmf}),
void abundances are increased or reduced according to their fiducial value of $\nu$ when $\sigma$ changes. The turnover radius $R_{\rm to}(z)$ marks the transition between increase and reduction. A negative shift $-\Delta \sigma$, due to e.g. increased neutrino suppression, produces a reduction of small/common voids ($\nu < 1$) and an increase of large/rare voids ($\nu > 1$). Note that the value of $\nu$ depends on the galaxy bias $b_{\rm g}$ through $\delta^{\rm v}_{\rm lin,m}$. 

The most significant effect on deep voids is straightforward. These voids are all larger than the turnover radius, and the suppression of the power spectrum increases the values of $\nu$ for the voids. Therefore, by Eq.~(\ref{eq:dlnfv}), the void abundances are reduced. 

The most significant effects on shallow voids are complicated by the fact that the turnover radius lies within the survey. This is because shallow voids are close to linear with almost equal Eulerian and Lagrangian radii. Some smaller voids here lie below the turnover radius, and have $\nu < 1$. The abundance of such voids is increased when the power spectrum is suppressed, by Eq.~(\ref{eq:dlnfv}). Above the turnover radius, however, the abundance of larger voids is instead reduced as a consequence of the power-spectrum suppression (as for deep voids). On large scales, $k \lesssim k_{\rm eq}$, the effective enhancement of the matter power spectrum (due to keeping $\sigma_8$ fixed, as discussed above) means that the abundance of the largest voids is instead increased.

\subsection{Forecast neutrino hierarchy constraints}
In Table~\ref{tab:constr}, we report forecast values for the Bayes factor (odds) $B_{{\rm NO,IO}}$, its logarithm $\ln \left(B_{{\rm NO,IO}}\right)$, and the effective significance level $n_{\sigma}^{\rm eff}$ with which the inverted hierarchy can be rejected (all under the assumption of a fiducial minimal normal neutrino hierarchy). The EC cluster survey and EV-A void survey cannot distinguish between the two different neutrino hierarchies on their own. The EV-B void survey and the combined EV-B + EC surveys could provide strong or decisive evidence against the inverted hierarchy, while the combined EV-A + EC surveys would only provide weak evidence against the inverted hierarchy (in the language of the conventional Jeffreys scale \cite{1939thpr.book.....J}). 

Like for the forecast parameter constraints, we also make a very conservative estimate of the potential degradation of our results due to cluster-void correlations, by completely throwing away the clusters in the overlapping volume across $z=0.7-2.0$ (EC-LO). We find that the above conclusions are robust with respect to cluster-void correlations.

For the alternative fiducial model, in which we set $h~=~0.7, \Omega_{\rm m}=0.3, \sigma_8=0.8$, the evidence against the inverted neutrino hierarchy is only marginally weaker than for the {\it Planck} 2015 cosmology. Conclusions remain unchanged. 

\subsection{Systematics}
The void abundance model is approximate, and a phenomenological extension of fits to $N$-body simulations. To account for observational and theoretical uncertainty, we include statistical scatter in void radius measurements (and likewise for cluster mass) and allow the characteristic void density contrast to vary through the parameter $D_{\rm v}$ in the analysis. We see no significant difference in the constraints compared to keeping the value of $D_{\rm v}$ fixed. This is in line with earlier findings \cite{2015arXiv150307690P}. Also, recent studies indicate that the tracer bias remains simple within voids \cite{2017MNRAS.469..787P,2018arXiv180606860P}. This suggests that bias-weighted modeling of void abundances of different depth as used here can be expected to work fairly well. Nevertheless, the relation between galaxy bias of voids and the large-scale galaxy bias of a survey is in general scale/density-dependent. Neutrinos also induce a weak scale dependence in the bias \cite{2014PhRvD..90h3530L,2016JCAP...11..015B}. We expect that the impact of these features is relatively small and within the uncertainty included on $D_{\rm v}$, but these aspects (and the impact of full halo occupation statistics) require further simulation studies and method development. A promising approach is the ``cleaning method'' of Ronconi and Marulli \cite{2017A&A...607A..24R}. Void selection and other potential sources of observational bias also require more detailed studies.

We neglect the impact of cluster-void correlations in our main analysis, but find that even when discarding all clusters in the overlapping volume between {\it Euclid} clusters and voids (EC-LO), the combined cluster + void neutrino-mass constraints are degraded only marginally. Dark energy constraints are degraded by at most a factor of a few (though in an exact treatment presumably less, since the full cluster-only constraints EC are significantly tighter). 

There is additional cosmological information in void samples that we have not considered here. We have limited the analysis to $z = 0.7 - 2$ for {\it Euclid}, for which spectroscopy is expected and hence redshift-space systematics should be minimal. The sample could be extended down to $z=0.2$ using photometry. The ellipticity distribution of voids is sensitive to cosmological parameters (including neutrino mass) \cite{2015arXiv150603088M}. The shapes and dynamics of voids are also independently sensitive to the expansion history and growth rate \cite{2016PhRvL.117i1302H}, providing further prospects for limiting the effects of systematics. The combination of deep + medium + shallow void counts should further help calibrate void systematics \cite{2018PhRvD..97j3504S}.

We argue therefore that the EV-A and EV-B cases can be regarded as worst-case and best-case scenarios for voids. The cluster constraints can be considered a best-case scenario, though imperfect knowledge of cluster mass--observable scaling relations and other possible systematics are not expected to significantly degrade constraints \cite{2015arXiv150502165S}. The combination of clusters and voids will also help to minimize the impact of systematics. We leave the detailed impact of theoretical and observational systematics to be treated in future work.

\subsection{Comparison to other probes}
Current best limits on the sum of neutrino masses are $M_{\nu}~<~0.10$--$0.15$ eV (95 \% CL) assuming a flat $\Lambda$CDM cosmology \cite{2017PhRvD..96l3503V,2018arXiv180706209P}. The stated range reflects different assumptions about the Hubble parameter and {\it Planck} cosmic microwave background (CMB) systematics. Cosmological data weakly favours a normal mass hierarchy, with the level of significance depending on assumptions about parameter priors \cite{2018JCAP...03..011G}. Neutrino oscillation experiments provide stronger, though not yet decisive, evidence in favour of a normal hierarchy \cite{2017PhRvD..96l3503V,deSalas:2017kay,2018FrASS...5...36M}.

Cosmological data are likely to remain a very competitive, albeit model-dependent, probe of neutrino masses for the foreseeable future \cite{2015ARNPS..6559001P, 2018FrASS...5...36M}. The combination of CMB data from {\it Planck}, the proposed CORE or CMB-S4 missions, with future large-scale galaxy clustering, cosmic shear and intensity mapping surveys (e.g. {\it Euclid}, the Large Synoptic Survey Telescope, the Square Kilometer Array) are expected to reach $\sigma(M_\nu)\sim 15$--$25$ meV, $\sigma(w_0) \sim 0.002$--$0.008$ for minimal neutrino masses. For the ($w_0,w_a$) dark energy model, the expectations are $\sigma(M_\nu)\sim 15$--$30$ meV, $\sigma(w_0) \sim 0.002$--$0.02$, $\sigma(w_a) \sim 0.01$--$0.05$ \cite{2016arXiv161002743A, 2018JCAP...03..003R, 2018arXiv180108331S, 2018FrASS...5...36M}. Since knowledge of the optical depth to reionization $\tau$ is a limiting factor for CMB neutrino mass constraints, the addition of a prior on $\tau$ from epoch of reionization (EoR) 21cm data can further reduce these uncertainties \cite{2017JCAP...02..052A}. The 21cm power spectrum in the EoR (and earlier times) can also provide independent constraints on neutrino masses \cite{2018FrASS...5...36M}. CMB-S4 cluster abundance also has great potential to provide strong constraints on dark energy and neutrino mass \cite{2016arXiv161002743A}. 
  
Cluster and void number densities as a function of characteristic size $R_{\rm L}$ scale as $n(R_{\rm L}) \sim \exp(-\delta_{\rm c}^2/\sigma_{R_{\rm L}}^2)$, where $\delta_{\rm c}$ is some characteristic linear density contrast threshold and $\sigma_{R_{\rm L}}^2$ is the matter power spectrum smoothed on the scale $R_{\rm L}$. Hence, cluster and void number counts are exponentially sensitive to the matter power spectrum on different scales.  
In contrast to CMB experiments, cluster - void samples have the advantages of probing the redshift evolution of expansion and structure growth directly, and not being degenerate with the optical depth to reionization. 

 Comparing the expected measurement precision from future experiments discussed above to Table~\ref{tab:constr}, large-area cluster and void cosmology with, e.g., {\it Euclid} can clearly be competitive with these experiments, in the best-case EV-B scenarios. The worst-case EV-A scenarios are not as competitive, though still provide informative independent constraints. These conclusions are robust if also accounting for cluster-void correlations (EC-LO).

\section{CONCLUSION}
Combining void and cluster counts could enable strong, simultaneous constraints on dark energy properties and neutrino properties, competitive with other future survey probes. Voids drive the neutrino mass constraint thanks to their wide range of sensitivity to the matter power spectrum across scales. Clusters add an orthogonal sensitivity to the mean matter density $\Omega_{\rm m}$ that breaks degeneracy between expansion/growth and power spectrum to constrain both dark energy and neutrino mass.

Independent of other data and assuming minimal, normal-hierarchy neutrino masses, we forecast that {\it Euclid} joint cluster and void number counts could reach 
\begin{eqnarray}
    \sigma(M_\nu) & \lesssim & 15\,{\rm meV}\,, \\
    \sigma(w_0) & \lesssim & 0.02\,, \\ 
    \sigma(w_a) & \lesssim & 0.07\,, \\ 
    B_{\rm NO,IO} & \gtrsim & 130 : 1\,, \\
    {\rm FoM}(w_0, w_a) & \gtrsim & 4600\,,
\end{eqnarray}
if all information in void radius and density bins can be used (EV-B + EC-LO), or at worst
\begin{eqnarray}
    \sigma(M_\nu) & \lesssim & 52\,{\rm meV} ,\footnotemark \\
    \sigma(w_0) & \lesssim & 0.03\,, \\ 
    \sigma(w_a) & \lesssim & 0.12\,, \\ 
    B_{\rm NO,IO} & \gtrsim & 2.3 : 1\,, \\
    {\rm FoM}(w_0, w_a) & \gtrsim & 640\,,
\end{eqnarray}
\footnotetext{Corresponding to $M_\nu < 0.19$ eV $(\gtrsim 95\%)$, taking $M_\nu \geq 0$ into account.}if only the total number of deep voids above the survey limiting radius in redshift bins can be used (EV-A + EC-LO). Note however that the full cluster-only constraints (EC) on $(w_0, w_a)$ are stronger than these [$\sigma(w_0)~=~0.01, \sigma(w_a)~=~0.04, {\rm FoM}(w_0, w_a)=3500$], since EC-LO includes only low-redshift ($z=0.2-0.7$) clusters to account in a very conservative way for the potential degradation due to cluster-void correlations.

An inverted neutrino hierarchy could, in the best-case scenario, be rejected at the $\gtrsim 99\%$ level using Bayes factor model comparison (with uniform nonrestrictive parameter priors). In the worst-case scenario, it is not possible to statistically distinguish between the two neutrino mass hierarchies. These findings are robust with respect to cluster-void correlations and our alternative fiducial model (and thus robust to whether {\it Planck} 2015 or 2018 fiducial parameters are considered).

Since the combination of clusters and voids breaks parameter degeneracies in the histories of expansion and of structure growth, these conclusions are independent of whether dark energy has a constant, $w(z)=w_0$, or time-varying, $w(z)=w_0 + w_az/(1+z)$, equation of state. The most optimistic precision achievable with clusters and voids alone is $\sigma(M_\nu) \lesssim 11$ meV, $\sigma(w_0) \lesssim 0.009$, $\sigma(w_a) \lesssim 0.03$.

Cluster and void cosmology with future large-area surveys such as {\it Euclid} has the potential to provide competitive constraints on extended cosmological models including massive neutrinos or time-varying dark energy, in a way that is independent of the cosmic microwave background and the conventional probes in galaxy surveys (e.g. galaxy clustering, weak lensing / cosmic shear). The development of the theoretical aspects and of the data analysis methodology which will allow us to fully exploit the potential of cluster and void counts will be the subject of further studies.

\section{Acknowledgments}
Thanks are due to O.~Botner, M.~Gerbino, A.~Hawken, A.~Pisani, C.~Reyes de los Heros, S.~Riemer-S{\o}rensen, R.~Sheth, J.~Silk and F.~Villaescusa-Navarro for useful comments and conversations. I am grateful to the anonymous referee for very constructive feedback. The computations were performed on resources provided by SNIC through the Uppsala Multidisciplinary Center for Advanced Computational Science (UPPMAX) under Projects No. SNIC 2017/1-260 and No. SNIC 2018/3-283. M.~S. was supported by Stiftelsen Olle Engkvist Byggm\"astare Project No. 2016/150.

\end{document}